\documentstyle[12pt]{article}

\catcode`\@=11
\long\def\@makefntext#1{ 
\protect\noindent \hbox to 3.2pt {\hskip-.9pt
$^{{\ninerm\@thefnmark}}$\hfil}#1\hfill} 

\def\thefootnote{\fnsymbol{footnote}}
 \def\@makefnmark{\hbox to 0pt{$^{\@thefnmark}$\hss}}  

\def\ps@myheadings{\let\@mkboth\@gobbletwo
\def\@oddhead{\hbox{} 
\rightmark\hfil\ninerm\thepage}
\def\@oddfoot{}\def\@evenhead{\ninerm\thepage\hfil 
\leftmark\hbox{}}\def\@evenfoot{}
\def\sectionmark##1{}\def\subsectionmark##1{}}

\begin{document}

\newcommand{\symbolfootnote}{\renewcommand{\thefootnote}
	{\fnsymbol{footnote}}}
\renewcommand{\thefootnote}{\fnsymbol{footnote}}
\newcommand{\alphfootnote}
	{\setcounter{footnote}{0}
	 \renewcommand{\thefootnote}{\sevenrm\alph{footnote}}}

\newcounter{sectionc}\newcounter{subsectionc}\newcounter{subsubsectionc}
\renewcommand{\section}[1] {\vspace{0.6cm}\addtocounter{sectionc}{1}
\setcounter{subsectionc}{0}\setcounter{subsubsectionc}{0}\noindent
	{\bf\thesectionc. #1}\par\vspace{0.4cm}}
\renewcommand{\subsection}[1] {\vspace{0.6cm}\addtocounter{subsectionc}{1}
	\setcounter{subsubsectionc}{0}\noindent
	{\it\thesectionc.\thesubsectionc. #1}\par\vspace{0.4cm}}
\renewcommand{\subsubsection}[1]
{\vspace{0.6cm}\addtocounter{subsubsectionc}{1}
	\noindent {\rm\thesectionc.\thesubsectionc.\thesubsubsectionc.
	#1}\par\vspace{0.4cm}}
\newcommand{\nonumsection}[1] {\vspace{0.6cm}\noindent{\bf #1}
	\par\vspace{0.4cm}}

\newcounter{appendixc}
\newcounter{subappendixc}[appendixc]
\newcounter{subsubappendixc}[subappendixc]
\renewcommand{\thesubappendixc}{\Alph{appendixc}.\arabic{subappendixc}}
\renewcommand{\thesubsubappendixc}
	{\Alph{appendixc}.\arabic{subappendixc}.\arabic{subsubappendixc}}

\renewcommand{\appendix}[1] {\vspace{0.6cm}
        \refstepcounter{appendixc}
        \setcounter{figure}{0}
        \setcounter{table}{0}
        \setcounter{equation}{0}
        \renewcommand{\thefigure}{\Alph{appendixc}.\arabic{figure}}
        \renewcommand{\thetable}{\Alph{appendixc}.\arabic{table}}
        \renewcommand{\theappendixc}{\Alph{appendixc}}
        \renewcommand{\theequation}{\Alph{appendixc}.\arabic{equation}}
        \noindent{\bf Appendix \theappendixc #1}\par\vspace{0.4cm}}
\newcommand{\subappendix}[1] {\vspace{0.6cm}
        \refstepcounter{subappendixc}
        \noindent{\bf Appendix \thesubappendixc. #1}\par\vspace{0.4cm}}
\newcommand{\subsubappendix}[1] {\vspace{0.6cm}
        \refstepcounter{subsubappendixc}
        \noindent{\it Appendix \thesubsubappendixc. #1}
	\par\vspace{0.4cm}}

\def\abstracts#1{{
	\centering{\begin{minipage}{30pc}\tenrm\baselineskip=12pt\noindent
	\centerline{\tenrm ABSTRACT}\vspace{0.3cm}
	\parindent=0pt #1
	\end{minipage} }\par}}

\newcommand{\bibit}{\it}
\newcommand{\bibbf}{\bf}
\renewenvironment{thebibliography}[1]
	{\begin{list}{\arabic{enumi}.}
	{\usecounter{enumi}\setlength{\parsep}{0pt}
\setlength{\leftmargin 1.25cm}{\rightmargin 0pt}
	 \setlength{\itemsep}{0pt} \settowidth
	{\labelwidth}{#1.}\sloppy}}{\end{list}}

\topsep=0in\parsep=0in\itemsep=0in
\parindent=1.5pc

\newcounter{itemlistc}
\newcounter{romanlistc}
\newcounter{alphlistc}
\newcounter{arabiclistc}
\newenvironment{itemlist}
    	{\setcounter{itemlistc}{0}
	 \begin{list}{$\bullet$}
	{\usecounter{itemlistc}
	 \setlength{\parsep}{0pt}
	 \setlength{\itemsep}{0pt}}}{\end{list}}

\newenvironment{romanlist}
	{\setcounter{romanlistc}{0}
	 \begin{list}{$($\roman{romanlistc}$)$}
	{\usecounter{romanlistc}
	 \setlength{\parsep}{0pt}
	 \setlength{\itemsep}{0pt}}}{\end{list}}

\newenvironment{alphlist}
	{\setcounter{alphlistc}{0}
	 \begin{list}{$($\alph{alphlistc}$)$}
	{\usecounter{alphlistc}
	 \setlength{\parsep}{0pt}
	 \setlength{\itemsep}{0pt}}}{\end{list}}

\newenvironment{arabiclist}
	{\setcounter{arabiclistc}{0}
	 \begin{list}{\arabic{arabiclistc}}
	{\usecounter{arabiclistc}
	 \setlength{\parsep}{0pt}
	 \setlength{\itemsep}{0pt}}}{\end{list}}

\newcommand{\fcaption}[1]{
        \refstepcounter{figure}
        \setbox\@tempboxa = \hbox{\tenrm Fig.~\thefigure. #1}
        \ifdim \wd\@tempboxa > 6in
           {\begin{center}
        \parbox{6in}{\tenrm\baselineskip=12pt Fig.~\thefigure. #1 }
            \end{center}}
        \else
             {\begin{center}
             {\tenrm Fig.~\thefigure. #1}
              \end{center}}
        \fi}

\newcommand{\tcaption}[1]{
        \refstepcounter{table}
        \setbox\@tempboxa = \hbox{\tenrm Table~\thetable. #1}
        \ifdim \wd\@tempboxa > 6in
           {\begin{center}
        \parbox{6in}{\tenrm\baselineskip=12pt Table~\thetable. #1 }
            \end{center}}
        \else
             {\begin{center}
             {\tenrm Table~\thetable. #1}
              \end{center}}
        \fi}

\def\@citex[#1]#2{\if@filesw\immediate\write\@auxout
	{\string\citation{#2}}\fi
\def\@citea{}\@cite{\@for\@citeb:=#2\do
	{\@citea\def\@citea{,}\@ifundefined
	{b@\@citeb}{{\bf ?}\@warning
	{Citation `\@citeb' on page \thepage \space undefined}}
	{\csname b@\@citeb\endcsname}}}{#1}}

\newif\if@cghi
\def\cite{\@cghitrue\@ifnextchar [{\@tempswatrue
	\@citex}{\@tempswafalse\@citex[]}}
\def\citelow{\@cghifalse\@ifnextchar [{\@tempswatrue
	\@citex}{\@tempswafalse\@citex[]}}
\def\@cite#1#2{{$\null^{#1}$\if@tempswa\typeout
	{IJCGA warning: optional citation argument
	ignored: `#2'} \fi}}
\newcommand{\citeup}{\cite}

\def\fnm#1{$^{\mbox{\scriptsize #1}}$}
\def\fnt#1#2{\footnotetext{\kern-.3em
	{$^{\mbox{\sevenrm #1}}$}{#2}}}

\font\twelvebf=cmbx10 scaled\magstep 1
\font\twelverm=cmr10 scaled\magstep 1
\font\twelveit=cmti10 scaled\magstep 1
\font\elevenbfit=cmbxti10 scaled\magstephalf
\font\elevenbf=cmbx10 scaled\magstephalf
\font\elevenrm=cmr10 scaled\magstephalf
\font\elevenit=cmti10 scaled\magstephalf
\font\bfit=cmbxti10
\font\tenbf=cmbx10
\font\tenrm=cmr10
\font\tenit=cmti10
\font\ninebf=cmbx9
\font\ninerm=cmr9
\font\nineit=cmti9
\font\eightbf=cmbx8
\font\eightrm=cmr8
\font\eightit=cmti8
\input epsf

\centerline{\tenbf RECENT DEVELOPMENTS IN WEAK LENSING}
\baselineskip=22pt
\vspace{0.8cm}
\centerline{\tenrm NICK KAISER$^1$, GORDON SQUIRES$^2$, GREG FAHLMAN$^3$}
\baselineskip=13pt
\centerline{\tenrm DAVID WOODS$^3$ and TOM BROADHURST$^4$}
\baselineskip=13pt
\vspace{0.8cm}
\centerline{\tenit $^1$CITA, University of Toronto, Toronto M5S 1A7, Canada}
\baselineskip=13pt
\centerline{\tenit $^2$Physics, University of Toronto, Toronto M5S 1A7, Canada}
\baselineskip=13pt
\centerline{\tenit $^3$Geophysics and Astronomy, UBC, Vancouver, Canada}
\baselineskip=13pt
\centerline{\tenit $^4$Astronomy, Johns Hopkins, Baltimore MD 21218}
\vspace{0.9cm}
\abstracts{Measurement of the gravitational distortion of images of
distant galaxies is rapidly becoming established as a powerful probe
of the dark mass distribution in clusters of galaxies. With the advent
of large mosaics of CCD's these methods should provide a composite total
mass profile for galaxy haloes and also measure the power spectrum of
mass fluctuations on supercluster scales.
We describe how HST observations have been used to place the
observational measurement of the shear on a quantitative footing.
By artifically stretching and then degrading WFPC2 images to simulate
ground based observing it is now possible to directly calibrate the effect
of atmospheric seeing.  Similar experiments show that one can
remove the effect of artificial image anisotropy arising in the
atmosphere or telescope.
There have also been important advances in the interpretation of the
shear: reconstruction techniques have been extended to encompass the
strong distortion regime of giant arcs etc., progress has been
made in removing a bias present in earlier reconstruction techniques,
and we describe new techniques for `aperture densitometry'.
We present some new results on clusters of galaxies, and
discuss the intimate connections between weak lensing and deep
spectroscopy.
}

\vfil
\rm\baselineskip=14pt
\section{Introduction}
The basic idea of weak lensing studies of clusters is well
illustrated by the crude simulation of figure \ref{fig:simulation}.
This shows a single `screen' of galaxy like blobs seen through
a softened isothermal sphere lens.  Near the centre one sees strong
distortion.  Less obvious to the eye, but easily detectable
by the computer, is the weak distortion at larger radii; the shear
falls as $1/r$ and is about 10\% at the edge of the frame.
While the shear falls off, the number of galaxies rises and therefore
so does the sensitivity, and in fact with this type of lens it is
in principle possible to map the cluster profile out to arbitrarily large
radii. At some point the halo of the target cluster will cease to dominate
the signal and at larger angular scales one will be measuring the stochastic
effect of large-scale structures along the line of sight adding in quadrature.

\begin{figure}
\centerline{\epsfxsize=400pt \epsfbox{sim.ps}}
\caption{Illustrating of the effect of a
isothermal sphere cluster lens model on a crude simulation
of a `source plane' covered with galaxy like blobs.}
\label{fig:simulation}
\end{figure}

Locally the distortion of the image of a galaxy
is a simple Lagrangian mapping of the
surface brightness:
\begin{equation}
\label{eq:lagrangianmapping}
f'(\theta_i) = f(\psi_{ij} \theta_j)
\end{equation}
For weak lensing the distortion tensor $\psi$ is close to the unit matrix and
for a lens at a single redshift, a good approximation for massive cluster
lenses which entirely dominate over foreground and background
`clutter', $\psi_{ij} = \delta_{ij} - \phi_{,ij}$ where $\phi$ is the
surface potential which is related to the surface density $\kappa$
by poisson's equation in 2-dimensions.

In order to reconstruct the mass of the lens in figure \ref{fig:simulation}
one need only locate the galaxies, measure their shapes and infer
at each point some locally averaged measure of the distortion (this
is described, e.g.~by the ratio and orientation of the eigenvalues
of the shear matix; or equivalently the axis ratio and position angle
for an intrinsically circular object). Note that as
real galaxies have substantial random intrinsic ellipticities
any weak distortion measurement is necessarily statistical in nature.
In the weak regime the
distortion is just equal to the shear $\gamma = (\phi_{11} - \phi_{22})/2$,
and from the shear field one can then reconstruct the surface
density (up to an unknown additive constant).

In reality of course things are quite complicated and observationally one must
worry about such things as contamination by cluster galaxies;
the effect of noise and seeing and anisotropy of the instrumental psf.
In interpreting the distortion one needs to know the mean redshift
distribution of the background galaxies, and one might worry
about such factors as the possibility of correlated intrinsic ellipticities
and fluctuations in the redshifts to the background galaxies due to
the `picket fence' structure. Here we will review some recent efforts to
address
some of these issues.

\section{Measuring the Distortion}

Several groups have developed software to detect and measure shapes
of faint galaxies.  A common feature of the shape analyses is the
use of second central moments $Q_{ij} \sim \int d^2 r\;r_i r_j f(\theta)$
to detect the statistical anisotropy, though
there are subtle differences in how these are calculated.  Some
kind of radial cut-off is needed to keep the noise finite, but this can be done
in a number of ways.  We use a gaussian
profile weight function with scale matched to the size of the object
as determined by our object detection algorithm\cite{KSB}, but
an alternative would be to use an isophotal limit
as in FOCAS\cite{FOCAS}. There
are then many different ways one can characterise the anisotropy.
An interesting
feature of these studies is that a general distortion pattern has
two real degrees of freedom --- the orientation and ratio of the $\psi_{ij}$
eigenvalues say --- but a real gravitational shear is derived from a
single scalar function $\phi$, so true gravitational shear patterns form
a zero measure subset of all possible distortion patterns.  This means
one could in principle use only the orientation of the eigenvalues
and discard entirely the information in the eigenvalue ratio for example,
and still reconstruct the surface density.  We do not recommend this,
but it illustrates the enormous freedom of approach here.  We follow Tyson's
approach\cite{TVW} and form a two-component `polarisation'
$e_1 = (Q_{11}-Q_{22}) / T$,
$e_2 = 2 Q_{12}/T$, with $T = Q_{11}+Q_{22}$ the trace of the quadrupole
tensor, and a similar statistic has been used by
Bonnet and Mellier\cite{BM94} though they normalise to $\sqrt{\det Q}$ rather
than to the trace.

\begin{figure}[t]
\centerline{
\epsfxsize=256pt \epsfbox{polarisation.ps}
}
\caption{Polarisation values
for a set of gaussian ellipsoid objects of various ellipticities and
orientation.}
\label{fig:polarisation}
\end{figure}

The polarisation is essentially a measure of the ellipticity (see figure
\ref{fig:polarisation}).  The nice thing about this statistic is that
it should average to zero for an unlensed population, and the shift
in the mean polarisation due to lensing is proportional to the shear
for weak distortions.

\subsection{Quantitative Calibration}
\vspace*{-0.35cm}
The calibration of the polarisation-shear
relation is a critical issue.  If one ignores the effect of
seeing and noise then one can perform a linearised analysis\cite{KSB}
to show that
the shift in the polarisation of an object due to a small perturbing
shear is
\begin{equation}
\label{eq:shearpolarisability}
\delta e_\alpha = P^\gamma_{\alpha\beta} \gamma_\beta
\end{equation}
where the shear polarisability tensor $P^\gamma_{\alpha\beta}$
is some messy, but
observable, combination of moments of the image brightness, so
an estimate of the shear is
\begin{equation}
\label{eq:shearestimate}
\delta \hat \gamma_\alpha = (P^\gamma_{\alpha\beta})^{-1} e_\beta
\end{equation}
The problem, for ground based observation at least,
is that atmospheric seeing will dilute the polarisation values
so (\ref{eq:shearestimate}) will underestimate the true shear for
small objects and we need some way to empirically calibrate this.
One approach\cite{FKSW} is to artificially stretch ground based images,
convolve them with a psf say twice the size of the real psf and
then see how the shear estimates are diluted as a function of
the galaxy size relative to the artificial seeing radius.  This is not
ideal as it assumes that faint galaxies are just scaled down
replicas of their brighter, and better resolved, cousins.

\begin{figure}[t]
\centerline{
\epsfxsize=350pt \epsfbox{seeingon.ps}
}
\caption{Result of an experiment to use HST data to calibrate the
polarization-shear relation in the presence of atmospheric
seeing as described in the text.}
\label{fig:calibration}
\end{figure}

With the HST fixed it is now possible to perform a more
direct calibration of the shear-polarisation relation\cite{KSB}.
Deep I-band WFPC2 images of random fields were artificially
stretched by an amount corresponding to a shear of 15\%
and then smeared with an artificial psf, rebinned
from $0.1''$ to $0.2''$ pixels, and had noise added
to simulate ground based observations.  These images were
then analysed in exactly the same way as the ground based data
and the results are shown in figure
\ref{fig:calibration}.
The systematic shift of the shear estimates is clearly seen
in the upper left panel, and the upper right panel shows
$\delta\gamma_\alpha$, the
change in the $\gamma_\alpha$ estimates induced by switching
on the shear.
In the lower panels $\delta\gamma$ is plotted
as a function of galaxy size and magnitude.  As expected,
$\hat \gamma_\alpha$ correctly measures the shear for the larger
objects but for the small objects we only recover a diluted signal.
With more data it should be possible to develop a detailed
model for the dilution factor as a function of image size, magnitude
etc.~and in different passbands which can be used to calibrate
future observations.

\subsection{Correction for psf Anisotropy}
\vspace*{-0.35cm}

\begin{figure}[t]
\centerline{
\epsfxsize=256pt \epsfbox{smearbefore.ps}
\epsfxsize=256pt \epsfbox{smearafter.ps}}
\caption{The four panels an the left show the polarisation values for
galaxies smeared with an anisotropic psf to mimic atmospheric
and instrumental effects.  At top left are the polarisation estimates,
and immediately to the right are the changes in the polarisation
$\delta e_\alpha$ caused by the psf anisotropy; this removes the random
scatter from the instrinsic ellipticities and shows the systematic
shift more clearly.  Below we show how the
polarisation shift depends on image size and magnitude.
The four panels on the right show the same quantities
values after we have selected a subsample of stars, measured
polarisations, fit for $p_\alpha$ and then corrected
the galaxy polarisation values. Clearly, this has efficiently
removed the effect of the instrumental psf anisotropy.}
\label{fig:hstsmear}
\end{figure}

Another critical issue is that of correction for anisotropy
of the instrumental point spread function (psf)
which can arise from a number of sources. Some, like
atmospheric dispersion, are predictable while others are of a stochastic
nature.  Luckily, most effects seem to produce a psf which varies
slowly across the image and, since one has a control sample
of typically 30 or so moderately bright foreground stars
per image one can map the psf and correct for this.
One can show\cite{KSB} that the response of the polarisation of an object
to a small psf anisotropy is
\begin{equation}
\label{eq:smearpolarisability}
\delta e_\alpha = P^s_{\alpha\beta} p_\beta
\end{equation}
where $p_\beta$ is a certain measure of the psf anisotropy
and $P^s_{\alpha\beta}$ is the `smear polarisability
tensor' which one can estimate for each object; it is essentially
just a measure of the inverse area of the object.
This allows one to infer $p_\beta$ from the
polarizations measured for foreground stars and one can then correct the galaxy
polarizations to what they would have been for an observer
with perfectly circular seeing.
The nice feature of the smear polarisability $P^s_{\alpha\beta}$ is that it
is calculated directly from the observed image and needs no correction
for seeing.

The analysis leading to (\ref{eq:smearpolarisability})
is somewhat idealised in that the surface brightness is
assumed to be a continuous function rather than a set of discrete
pixel values and it ignores noise which can bias the polarizability.
A rigorous test of this procedure is allowed by the HST
data used above.  Here we do not stretch the images but rather
smear them with an anisotropic psf, rebin them, add noise and
then analyse as before.  The results of this, are shown in
figure \ref{fig:hstsmear}.  Clearly the method has worked
very well.

\section{Interpreting the Distortion}
The experiments described above show that the gravitational shear can be
measured in a quantitative manner and that the problem
of instrumental psf anisotropy is well in hand.  Assuming one
is supplied with the shear estimates for the background
galaxies the next step is to recover the surface density
or surface potential responsible for this.  There have been
a number of recent advances in this which we now briefly
describe.

\subsection{Reconstruction Methods}
\vspace*{-0.35cm}
The reconstruction method of Kaiser and Squires\cite{KS93}
(hereafter KS93) suffers from at least
two shortcomings: it is limited to the weak distortion regime
and it suffers from a bias near the edge of the reconstruction.
We consider first the question of bias.  A useful way to
analyse the problem is to start with the local expression
for the gradient of the surface density in terms of gradient
of the shear\cite{KSFW}$^,$\cite{NK94}
\begin{equation}	\label{eq:gradkappa}
\vec\nabla\kappa =
\left[
\begin{array}{c}
\gamma_{1,1} + \gamma_{2,2} \\
\gamma_{2,1} - \gamma_{1,2}
\end{array}
\right]
\end{equation}
A direct consequence of this that one can in principle make any
{\sl differential\/} surface density measurement, by integrating the observable
$\nabla \kappa$ along a line, but there will always be
an ambiguity in the baseline. It suggests that one might
measure the surface density at some point relative to the
mean value on the boundary say by averaging over line integrals
of $\nabla \kappa$ from the point in question to the boundary
as indicated schematically in figure \ref{fig:finite}a.

\begin{figure}
\centerline{
\epsfxsize=150pt \epsfbox{pathsa.ps}
\epsfxsize=150pt \epsfbox{pathsb.ps}
\epsfxsize=150pt \epsfbox{pathsc.ps}}
\caption{Panel a shows schematically the line integrals
averaged over to measure $\kappa$ relative to the mean value
on the boundary as in in Schneider's method.  Panel b shows
an alternative approach in which one measures $\kappa$ relative
to the average over the area surveyed, and
panel c illustrates a hybrid scheme.  For noise free data
these methods are exactly equivalent, but case-b is simplest
to implement and seems to have better noise characteristics.}
\label{fig:finite}
\end{figure}

An ingeneous algorithm to implement such a scheme has been
developed by Schneider\cite{S94}.
His method works by smoothing the shear field, calculating the
gradient of the shear, performing the line integrals
numerically and then averaging.
The method appears to work
well with simulated data but it is not clear it is optimal.
An alternative would be to measure the surface density relative
to its value averaged over the area of the survey
as in figure \ref{fig:finite}b or relative to some other
reference region as in \ref{fig:finite}c.  Since one is always
free to adjust the baseline surface density after the fact
these methods are all, for noise free data, exactly equivalent.

Performing the line integrations by parts analytically\cite{KS94}
one can express the surface density for any method of this kind as
\begin{equation}
\kappa(\vec r_0) - \overline \kappa = {1\over A} \int d^2r
K_\alpha(\vec r; \vec r_0) \gamma_\alpha(\vec r)
\end{equation}
where $\overline \kappa$ is the reference value or baseline
surface density.  For case-b the kernel is simply
$K_\alpha = p^3 \gamma^*_\alpha / r^2 d$ where we have placed the
`target point' $\vec r_0$ at the origin of coordinates and where $p$, $d$
and the unit shear vector $\gamma^*_\alpha = -\{
\cos 2 \varphi + \theta, \sin 2 \varphi + \theta\}$ are as shown in
figure \ref{fig:angles}, so
\begin{equation}
\label{eq:ks94kappa}
\kappa - \overline \kappa = {1\over A} \int d^2r
{p^3 \over r^2 d} \gamma^*_\alpha \gamma_\alpha
\end{equation}
This is very similar to the KS93 estimator, and in fact they become identical
for target points close to the centre of a circular survey.
Both methods `project out' a particular component
of the shear; in the earlier method this was simply the tangential
shear, here the shear vector is rotated as shown in figure \ref{fig:angles}.
One can then construct an estimator as a simple discrete
sum over the individual galaxy shear estimates:
\begin{equation}
\label{eq:ks94estimator}
\hat \kappa = {1\over \overline n A} \sum
{p^3 \gamma^*_\alpha \hat \gamma_\alpha \over r^2 d}
\end{equation}
This is a little more complicated to
implement as one has to solve
for the boundary distance $p$ etc.~as a function of the measurement
point $\vec r$, but for simple survey geometries such as rectangles, circles
or ellipses this can easily be done, and (\ref{eq:ks94estimator})
provides a simple upgrade to KS93.  It avoids having to smooth
the shear field and requires no numerical differencing or
integration and has the
advantage that it is straightforward to calculate
the statistical uncertainty from the random
intrinsic ellipticities and photon counting
noise.\footnote{There is a more subtle problem with case-a
where one tries to pin $\overline \kappa$ to the
boundary.  The kernel then picks up a delta-function
term which measures a certain average of the shear on the
boundary.  This makes a significant contribution,
especially for the low spatial frequency components in the
reconstruction, but is tricky to estimate.  If the shear is smoothed, its value
on the boundary will be biased, and if not, the small number of galaxies
near the boundary give a large statistical error.}

\begin{figure}
\centerline{\epsfxsize=200pt \epsfbox{angles.ps}}
\caption{Graphical definition of the quantities $p$, $d$, and the
unit shear vector $\gamma_*$ appearing in the kernel.  The `target
point' $\vec r_0$ lies at the intersection of the cartesian
axes, the loop represents the boundary of the region observed,
and $p$ is the boundary point lying behind the `measurement point' $\vec r$.
The ellipse indicates the orientation of $\gamma^*$.}
\label{fig:angles}
\end{figure}

\subsection{Aperture Densitometry}
\vspace*{-0.35cm}
While reconstruction methods are very nice in that they provide
a direct 2-dimensional map of $\kappa$ with well defined
statistical uncertainty, sometimes one would
like to measure some gross statistic such as the mass within
a given aperture (perhaps after having located the cluster centre from
direct reconstruction methods).  Now one could simply perform densitometry
on the reconstruction, but then one would seem to face a non-trivial task in
calculating the statistical uncertainty.  In fact, one can simply modify
the estimator described above to do this: If we use
(\ref{eq:ks94estimator}) to calculate $\kappa -
\overline \kappa$ from the data within some boundary $p$ and then calculate
$\kappa -\overline \kappa'$ for some other boundary $p'$ enclosed
within $p$ then the difference is just $\overline \kappa -
\overline \kappa'$ the difference in the mean surface density for
the two nested apertures, and by simple algebra on can similarly
construct an estimate of $\overline \kappa$ within the inner aperture
relative to its mean value in the surrounding annulus.

Moreover, if one arranges that the aperture has the same shape as
the survey boundary, one finds that the two kernels
cancel exactly within the aperture and one obtains a rigorous
lower bound on the mass within the aperture using only data in the
surrounding control region.
This generalises a result previously
applied\cite{FKSW} with circular apertures, and is nice of one wants to avoid
contamination by cluster galaxies or if one wants to
be sure to remain in the weak lensing regime.  These aperture mass
estimates are made as a single discrete sum over the galaxy
ellipticities --- just a truncated version of (\ref{eq:ks94estimator}) ---
and have correspondingly simple statistical uncertainty.

\subsection{The Strong Distortion Regime}
\vspace*{-0.35cm}
There have also been advances in extending the reconstruction techniques
into the strong lensing regime.  While only relevant in the very centres
of clusters, these are the regions where one can hope to get the highest
spatial resolution image of the dark matter as the signal is
so strong.  In the weak regime the distortion directly measures the
shear.  In general the observable quantity ---
the ratio of the eigenvalues of the distortion tensor --- is determined
not by the shear alone, but\cite{GSF} by the combination
$\gamma / (1-\kappa)$.  Moreover, there
is an ambiguity in that the relation between these is double valued\cite{SS1}.
This is easily understood at a qualitative level:
As one approaches a critical line the distortion rises and the eigenvalue
ratio becomes infinite
as one of the eigenvalues vanishes.
As one crosses the line the parity flips and the distortion then decreases
again so two points, one either side of the critical line,
may have the same distortion but different values for
$\gamma / (1-\kappa)$ which varies smoothly and continuously across
the critical line.
It turns out that one can define an observable $e_\alpha$ which is
equal to $\gamma_\alpha / (1-\kappa)$ in the even parity regime, but in the
odd parity region is equal to the inverse of this vector.

One approach to this problem\cite{SS2} is to develop an iterative
scheme to solve this problem.  Quite possibly this will prove to be the
best approach.  There is however an alternative.  Just as one finds
in the weak regime a local relation between $\nabla \kappa$ and the
shear gradients, in the general case one can derive\cite{NK94} a
local expression for $\nabla \log (1-\kappa)$ in terms of the
observable $e_\alpha$, and from this one can construct a direct
reconstruction estimator for $\log (1-\kappa)$ in much the same
way as described above by averaging over line integrals.  There are
many practical problems to be overcome in applying this, but it
is clear from recent HST observations of the cores of very massive clusters
(Richard Ellis, personal communication) that
there is a profusion of arcs and the prospects seem quite good.

\section{New Results on Clusters}

We have reported elsewhere\cite{FKSW}$^,$\cite{KSFW} on MS1224,
and we give a partial update on this below.
We have subsequently obtained data on a number of other
clusters including A2218, A1689, A2390 and A2163.
Surface density reconstructions for A2218 and A1689 are shown in
figures \ref{fig:a1689},\ref{fig:a2218}.  These were made with
the KS93 estimator and so may suffer from a slight bias near the edge.
Also, these results have yet to be calibrated for the effect
of seeing; this will be done shortly using HST data as described
above.  Clearly though, these data are yielding a very strong
signal and we will hope soon to be able to present
quantitative mass estimates and mass-to-light
ratios for these clusters, as well as detailed shear profiles.

\begin{figure}
\centerline{\epsfxsize=400pt \epsfbox{A1689.ps}}
\caption{Mass reconstruction of A1689.  The contours show the
total mass density superposed on the V-band image. These data
were taken at the NTT.
The V-band image is approximately $8'$ on a side.}
\label{fig:a1689}
\end{figure}

\begin{figure}
\centerline{\epsfxsize=500pt \epsfbox{A2218.ps}}
\caption{Mass reconstruction of A2218.  As before, the contours show the
total mass density, now superposed on a mosaic of the I-band image. These data
were taken at the CFHT. The box here is approximately
$12'$ on a side.}
\label{fig:a2218}
\end{figure}

\begin{figure}
\centerline{\epsfxsize=400pt \epsfbox{beta.ps}}
\caption{Distribution of distortion strengths for a realistic distribution
of background galaxy redshifts and for various lens redshifts.
The lower panel shows (solid) the smoothed redshift distribution for the
I=20-22 CFRS redshift survey presented at this meeting which is nearly
complete and represents an increase of about an order of magnitude
over previous surveys at these magnitudes.
The dashed curve is an extrapolation to fainter magnitudes I=22-24 made
assuming
no evolution and kindly provided by Simon Lilly.
The upper panels show, for these two magnitude slices,
the distribution of  \protect{$\beta$} values --- the distortion strength
relative to
that for an infinitely distant
object --- for 5 lens redshifts $z_l = 0.1,0.15,0.2,0.3, 0.5$
progressing from right to left.
The curves move progressively to higher $\beta$ and become more sharply
peaked as the lens redshift decreases.}
\label{fig:beta}
\end{figure}

\section{Connection with Spectroscopy}

There is a very intimate connection between weak lensing and
spectroscopy, particularly the deeper surveys discussed
at this meeting.  The distortion observations allow one to
infer the dimensionless surface density $\kappa = \Sigma / \Sigma_{\rm crit}$.
In order to determine the physical surface density $\Sigma$ requires
knowledge of the background galaxy redshifts as this determines
$\Sigma_{\rm crit}$.  The key factor here is $\langle \beta \rangle
= \langle \max(0, 1-w_l/w_s) \rangle$ which gives the mean distortion
strength for a population of galaxies at finite distances $w_s$
relative to that for infinitely distant objects.
Unfortunately there are as yet no complete redshift
surveys which reach the faint magnitudes used here so some extrapolation
is needed.  This is illustrated in figure \ref{fig:beta}.
For lenses at moderately low redshifts, say $z \sim 0.2$ or less,
the distribution of distortion strengths is very sharp --- the `single
source screen' approximation is quite good, and the change in $\langle \beta
\rangle$ in going from I = 20-22, where the redshift distribtuion is
well established, to I = 22-24 say is quite small.
For such clusters our mass estimates are therefore quite insensitive to
uncertainty in the extrapolation of $n(z)$.

Our mass estimate\cite{FKSW}
for MS1224 ($z=0.33$) was made with a much smaller sample of redshifts and
using a
cruder extrapolation and we obtained $\langle \beta \rangle = 0.34$.
With the much larger sample here and using the `no-evolution' extrapolation
we obtain $\langle \beta \rangle = 0.38$, which reduces the mass
estimate by about 10\%. We also estimated $\Omega$ by dividing the
mass per galaxy for the cluster by the mass per (field) galaxy for a closed
universe (in the traditional manner of comparing apples with oranges) at
the cluster redshift.  The
result was $\Omega \sim 2$, much higher than the values traditionally obtained
for clusters. This was
in part because of the high mass-to-light
ratio we obtained and in part because of the apparently much higher
comoving number density of galaxies seen in these deep surveys
as compared with local studies.  This latter feature
seems to be supported by the newer CFRS results.  This is clearly of great
importance in interpreting any mass estimates, lensing or otherwise,
and needs to be better understood.

For higher redshift lenses the distortion
becomes weaker (for a fixed velocity dispersion lens), and the dependence
of $\langle \beta \rangle$ on redshift is much more marked.
One can turn this around and try
to determine the redshifts of faint galaxies by measuring the distortion
strength\cite{SMAIL}.
For a cluster lens at $z=0.5$, the mean distortion strengths
$\langle \beta \rangle$ for
the two redshift distributions shown in
figure \ref{fig:beta} differ by almost a factor 2.
It would be relatively easy to distinguish between these
two possibilities if the strength of the lens were
accurately known from velocity dispersion or X-ray temperatures,
but there may be some reason to doubt these. With only the lensing data
one can still measure the relative distortion as a function of magnitude.
On could imagine measuring the distortion for an I = 22-24 slice
relative to the I = 20-22 slice (for which $\langle \beta \rangle$ is
known).  The limiting factor here is measuring the rather weak distortion
for the bright sample for which $\langle \beta \rangle$ is only about 15\%,
and it will probably be necessary to stack the results from several clusters.
With lower redshift lenses things are no better. For $z_l = 0.3$ the mean
$\beta$ for the bright sample is $0.35$ but now the distortion strengths for
the two distributions shown in figure \ref{fig:beta} only differ
by about 25\%, so even distinguishing between these grossly different
redshift distributions would require several clusters to obtain the
necessary precision.
It will clearly be necessary to calibrate the shear measurement
very accurately, aspecially as the sizes of the galaxies, and therefore
the seeing dilution effect, will tend to
vary systematically with magnitude.  A robust way to do this would be
to observe a nearby low redshift cluster under the same conditions
since for such clusters the true distortion strength should
be very nearly independent of magnitude.
While technically challenging,
if one could firmly establish the relative mean distance to faint galaxies as
a function of magnitude from lensing then, when redshifts for such galaxies
become available from 10m class telescopes one will have a powerful
test of the cosmological world model.

\end{document}